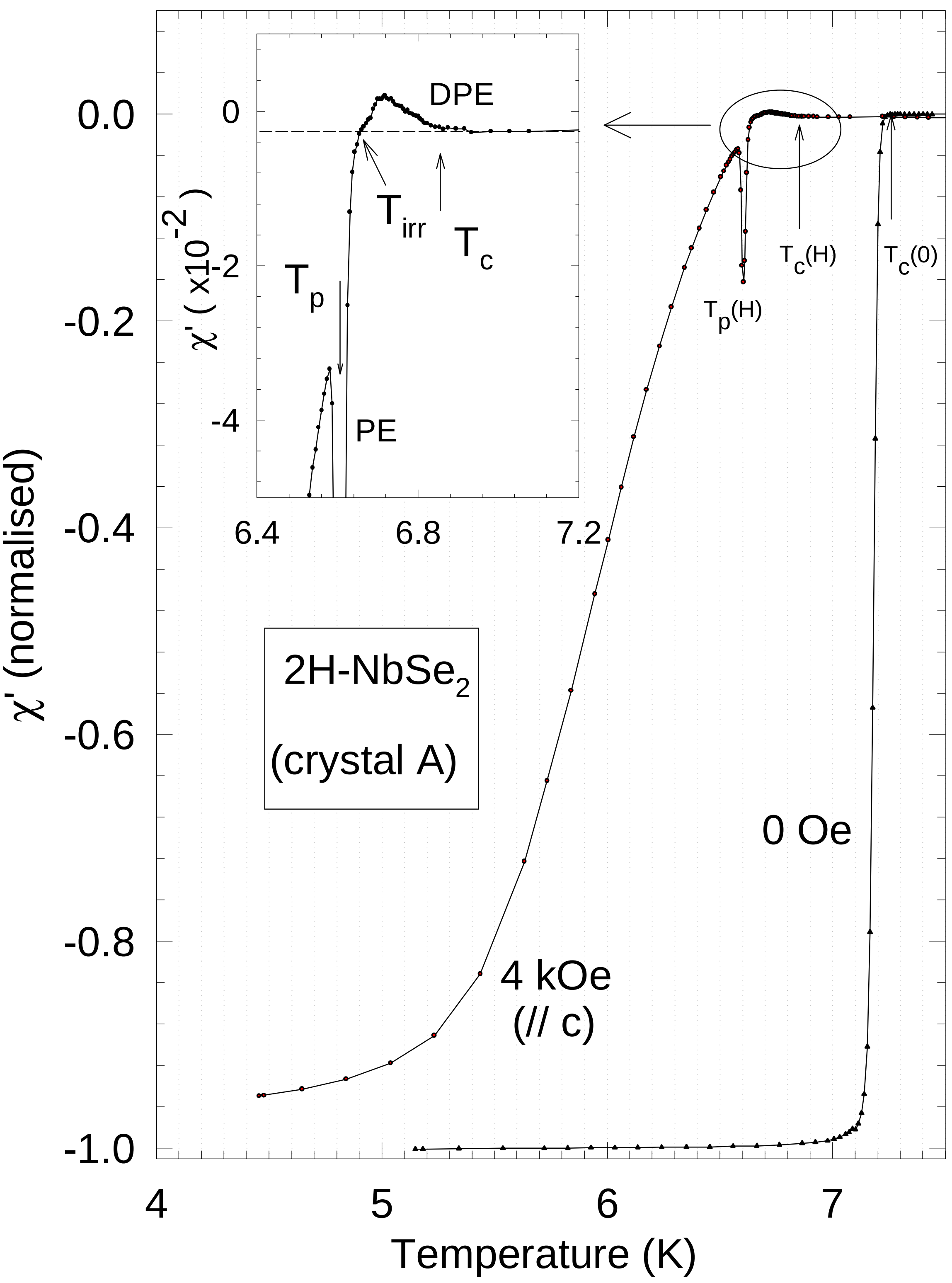

Fig : 1

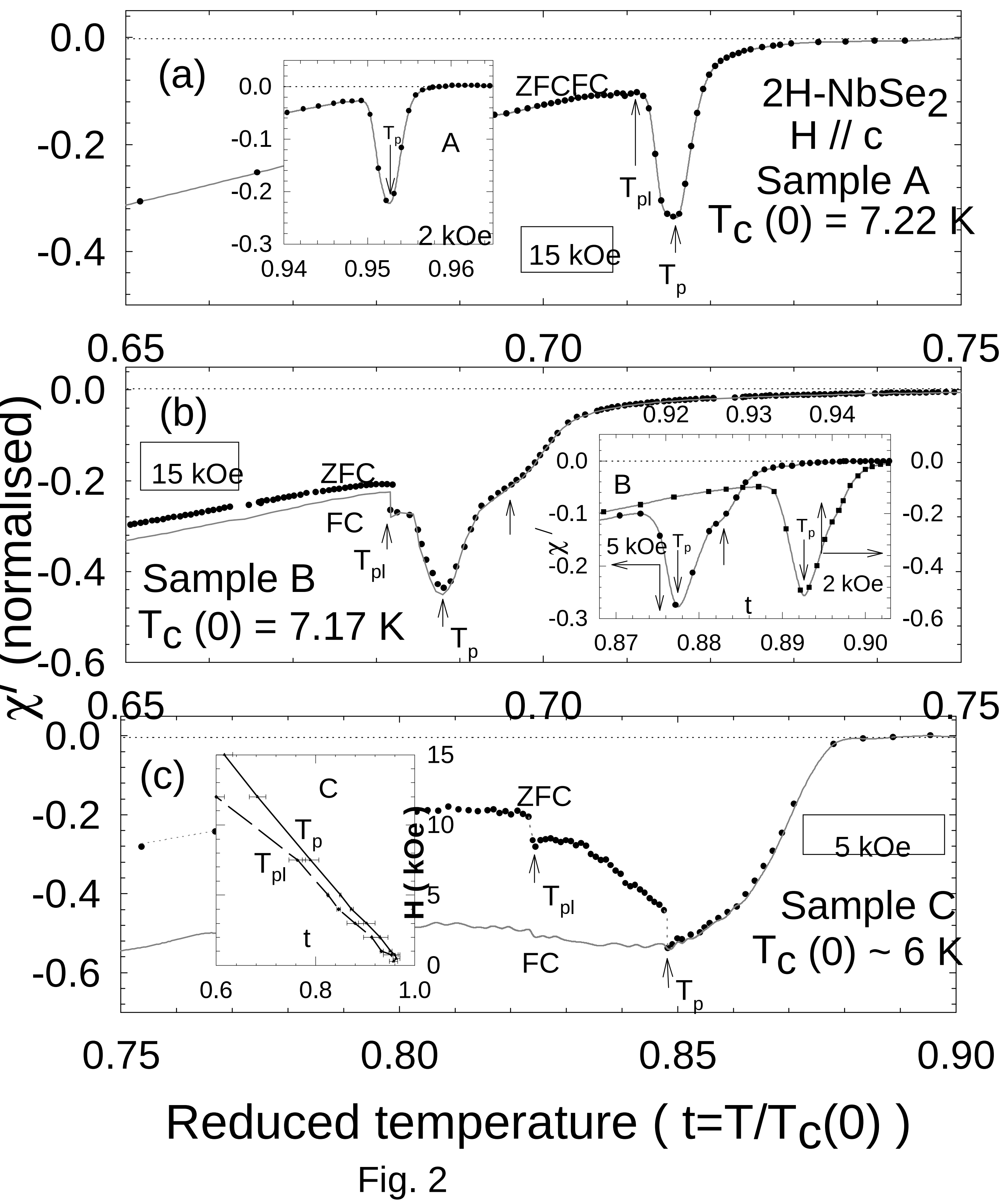

Fig. 2

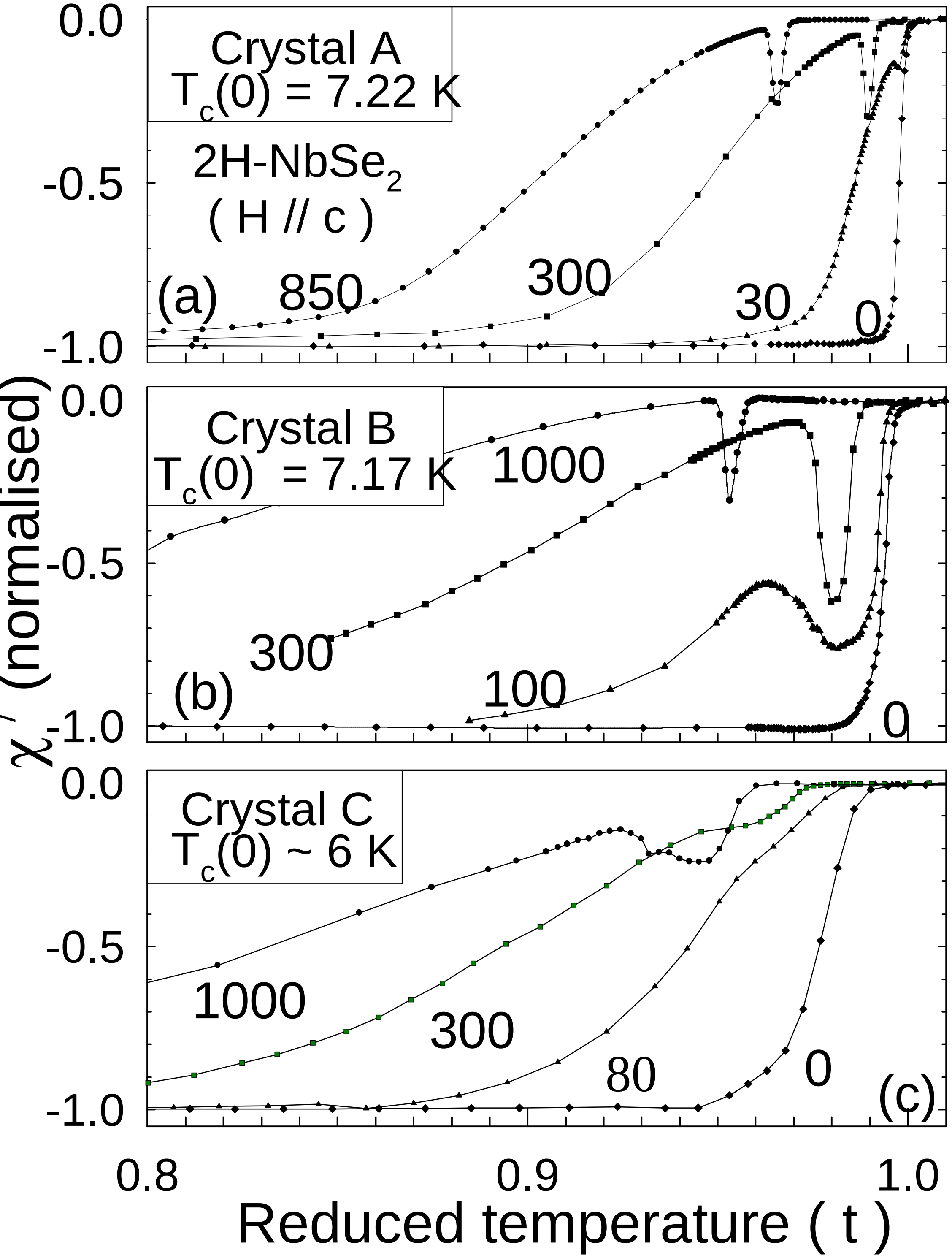

Fig. 3

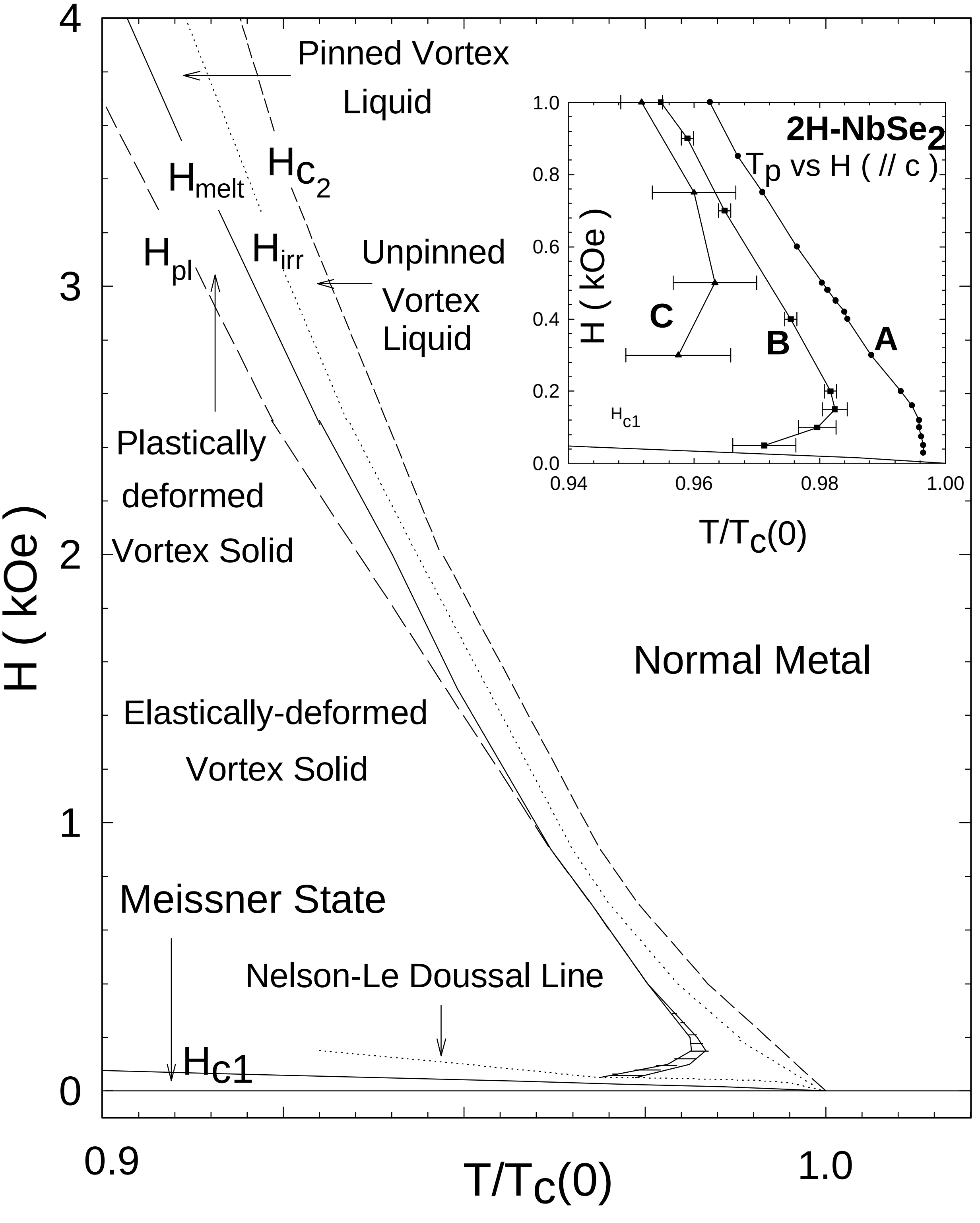

Fig.4

1 2 3 4 5 6 7 8 9 1011 12 13 14 15 16 17 18
, , , , , , , , , , , , , , , ,
19 20
,, , ,

# Disorder, History Dependence and Phase Transitions in a Magnetic Flux Line Lattice

Satyajit S. Banerjee[*], Subir Saha[*], Nitin G. Patil[*], Ramakrishnan S.[*], Arun K. Grover[*,@], Shobo Bhattacharya[*,#,@], Ravikumar G[+]., Prashanta K. Mishra[+], Chandrasekhar Rao T.V. [+], Vinod C. Sahni[+], Chakalakal V. Tomy[$,%], Geetha Balakrishnan[$], Don Mck. Paul[$] and Mark J. Higgins[#]

[*]Tata Institute of Fundamental Research, Mumbai-400005, India

[+]TPPED, Bhabha Atomic Research Centre, Mumbai-400085, India

[$]Department of Physics, University of Warwick, Coventry, CV4 7AL, U.K.

[#]NEC Research Institute, 4 Independence Way, Princeton, New Jersey 08540, USA

[%]Department of Physics, Indian Institute of Technology, Kanpur 208016, India.

**A magnetic flux line lattice or a vortex lattice, where each flux line contains a quantum of flux ($\Phi_0 = hc/2e$) and interacts repulsively, is expected to occur in the entire mixed state of a type-II superconductor, within a mean field description[1]. Inclusion of thermal fluctuations predicts melting of this vortex lattice into a vortex liquid state, for both dilute and concentrated arrays of lines, yielding an unusual reentrant phase boundary[2,3] in the magnetic field-temperature (H,T) plane, i.e., at a fixed T, the melting phase boundary is encountered twice. Quenched pinning disorder adds more variety in the form of novel disordered (glassy) phases[4,5], and makes the vortex lattice an ideal system to study the competition and interplay between interaction and disorder. Our experimental studies of magnetic response in the mixed state of the superconductor 2H-$NbSe_2$ (in samples of progressively increasing pinning) show the appearance and evolution of novel behavior such as history dependence, reminiscent of spin glasses[6]. These results elucidate how pinning alters the phase boundary separating the "ordered" and “disordered” states of the flux lines and causes a novel disorder-induced transition. A generic magnetic phase diagram of a type II superconductor, that includes effects of thermal and quenched disorder, has been constructed.**

All the data are obtained using the standard magnetic ac susceptibility studies involving the response to a small amplitude ac magnetic field ($h_{ac}$) superimposed on a background dc bias field ($H_{dc}$) maintained parallel to the c-axis of two layered hexagonal Niobium diselenide (2H-$NbSe_2$). The magnetic induction (B) determines the lattice constant $a_0 (= (2\Phi_0 / B\sqrt{3})^{1/2} \approx 5\times10^4 A^0 / \sqrt{B(gauss)})$ of the triangular flux line lattice (FLL)[5].

Fig.1 shows the typical variation of the real part of the ac susceptibility ($\chi^{/}$) with temperature (T) : it shows perfect diamagnetism at low T ($\chi^{/} \approx -1$) and rises to the normal state value at $T_c(0)$ in zero-field. For a dc bias field, $\chi^{/}$ can be approximated, in the Critical State model[7], by : $\chi^{/} \sim (-1 + \alpha h_{ac}/J_c)$, where $\alpha$ is a geometry dependent factor and $J_c$ is the critical current density at a given (H,T). With increasing T, $\chi^{/}$ monotonically increases due to an expected decrease in $J_c$. As shown in the curve for H = 4 kOe in Fig. 1, this usual behavior in $\chi^{/}$ is interrupted at the onset of an anomalous enhancement of diamagnetic screening response ; $\chi^{/}$ reaches a sharp minimum at $T_p$ ( which can be sharper than even the zero-field superconducting transition, as in Fig.1) , above which it recovers rapidly towards the normal state value at $T_c(H)$. This nonmonotonicity in $\chi^{/}$ is a consequence of a nonmonotonicity in $J_c$. The minimum in $\chi^{/}$ thus corresponds to a peak in $J_c$ , the ubiquitous peak effect[8] (PE). In weakly pinned systems, $J_c$ is given[5] by the pinning force ( $J_cB$ ) equation : $J_cB = (n_p\langle f_p^2\rangle/V_c)^{1/2}$, where $n_p$ is the density of pins, $f_p$ is the elementary pinning interaction, proportional to the condensation energy and $V_c$ is the correlation volume ( of a Larkin[4] domain within which flux lines retain their order) directly related to the elastic moduli of the FLL. The PE results[9] from a rapid collapse in $V_c$ of the Larkin domain ( more rapid than the decrease in $\langle f_p^2\rangle$, which causes the usual monotonic decrease in screening response with increasing temperature or field) so as to produce an overall enhancement in $J_c$ and hence in the diamagnetic response. This unusual collapse of $V_c$, i.e., the loss of order of the FLL, is triggered by an anomalously rapid softening of the FLL. Although precise quantitative explanations are still lacking, all available scenarios[10,11] intimately relate PE to the melting of the FLL. Indeed, the PE boundary in a few weak-pinning samples of 2H-$NbSe_2$ were found[12] to mimic closely the theoretically predicted[2] reentrant melting phase boundary of pinning free FLL. In what follows, we show in samples of varying quenched disorder (i.e., pinning centers) the evolution of the loss of order of FLL near the incipient melting transition. We focus on the temperature regime in between the onset of the peak effect and the reappearance, above $T_p$, of a conventional approach of $\chi^{/}$ to the normal state value ( hereafter referred to as the peak effect regime). Our results provide new insight on the details of the process of loss of order in the FLL and the appearance of phenomena such as history dependence ( and metastability ) and new pinning induced phase transitions.

I. *Disorder, Metastability and History dependence at moderate fields* :

Three different single crystals (samples A, B and C, made from starting materials of progressively decreasing purity), spanning two orders of magnitude differences in critical current densities ($J_c$ values in A to C range from few A/ $cm^2$ to few hundred A/ $cm^2$ at T=4.2K and in H ~ 10 kOe) were studied. In each of these crystals, the flux lattice is well formed : the transverse correlation length $R_c$[5] is much larger[9] than the values of the lattice constant $a_0$ at moderate to high fields ( H > 1 kOe ).

Fig.2 shows the results of a history-dependence experiment. Two different sample histories, commonly associated with disordered magnets such as spin-glasses, were investigated. In one, the sample is first heated to the normal state and cooled in the presence of a field (field-cooled, *i.e.*, FC). In the second, it is cooled in zero-field and then the field is applied (zero-field cooled, *i.e.*, ZFC). In both cases $\chi^{/}$ data were collected during the subsequent warm-up.

Fig.2(a) shows the temperature dependence of $\chi^{/}$ for two values of the field, 15 kOe and 2 kOe, in the cleanest sample A ( made from ultrapure starting materials, notably, nearly Ta-free Nb). There is no measurable history dependence; the ZFC and FC curves are the same, i.e., the flux lines system reaches the ordered state regardless of the path chosen. In Fig.2(b), the data from a typically clean crystal (sample B made from 99.9% pure starting materials) are shown for three values of the magnetic field. For H =15 kOe, the same field value as in Fig.2(a), history dependence appears : the FC state is slightly more diamagnetic at the lowest temperature. The onset of the peak effect regime occurs discontinuously through a downward jump in $\chi'$ at $T_{pl}$. The first peak in $\chi'$ at $T_p$ is followed by the appearance of a new peak (seen as a shoulder on the rapidly declining background response in $\chi'$). The two $\chi'$ curves, FC and ZFC, merge into one near this shoulder. For a lower field value of 5 kOe ( see inset in Fig. 2(b) ), both the features, the jump in $\chi'$ at the onset of peak regime and the history dependence in it, disappear as in Fig. 2(a), although the two peaks spanning the peak regime are still identifiable ( cf. arrows in the inset of Fig. 2(b) ). At a further lower field of 2 kOe, only one peak remains and the $\chi'$ curve now ( see inset in Fig. 2(b)) resembles the two curves in Fig.2(a). As stated above, at a given (H,T) value, the sample B has a larger $J_c$ than that in the sample A. The appearance of structure and history dependence in $\chi'$ ( in sample B ) in the peak effect regime thus appears to be a signature of enhancement of disorder / pinning. These results also suggest that in a given sample ( here, crystal B ), one produces an effectively stronger pinning, upon increasing H, as has been postulated earlier elsewhere[5,13].

In Fig.2(c), we show the data for our most strongly pinned FLL (in sample C, made from commercial grade starting material). In this sample, even for H=5 kOe, a large history dependence appears. ( Note the similarity between the results shown in Fig. 2(c) and those obtained near the spin glass transition temperature ($T_g$) in a canonical spin glass system, like, **Cu**Mn , in which ZFC and FC magnetization curves merge at $T_g$.[6]) The FC state in Fig. 2(c) is more diamagnetic, i.e., more strongly pinned than the ZFC. This is equivalent to the transport measurements[14] on another sample from the same batch of crystals of 2H-$NbSe_2$, which showed a far greater critical current density for the FC branch and which could be annealed ( by a driving current ) into a more ordered and more weakly pinned ( smaller $J_c$ ) state, like, the ZFC branch. The FC state in Fig. 2(c) results from the supercooling of the disordered vortex liquid state from higher temperatures; it nearly retains frozen-in liquid correlations and thus shows minimal signatures of any further loss of order for FLL on warming up across the peak regime. For the sample C, at lower fields ( < 2 kOe, data not shown here ), where pinning is effectively reduced, the situation is similar to that seen in sample B at 15 kOe. The history dependence and metastability are thus unambiguously related to increased pinning.

The ZFC branch in sample C shows a very broad peak effect regime ( see Fig. 2(c) ). However, unlike results in sample B ( cf. Fig. 2(b) ), a clear multi-peak structure is unidentifiable in sample C. Instead, the onset of peak effect regime is through a large jump in $\chi'$ at $T_{pl}$, and ending also with a large jump at $T_p$, above which much of the history dependence in $\chi'$ disappears. The sizes of the two jumps are amplitude dependent; they become larger with lower ac amplitude, implying that they are of an equilibrium origin, i.e., they mark two (discontinuous) phase transitions. Thus, with increased effective pinning, the PE feature of the flux lattice system evolves from a single history-independent and sharp peak ( in sample A ) to a strongly history dependent broad peak effect region ( in sample C ), which is spanned by two discontinuous transitions for the more ordered ZFC branch, but is nearly featureless, without a conspicuous PE, for the more disordered FC branch.

We propose that $T_{pl}$ marks a transition from a nearly defect-free lattice[15], such as a Bragg glass[16], to a highly defective plastically deformed lattice[16,17,18], full of topological defects (dislocations) analogous to a "vortex glass" phase proposed by Gingras and Huse[19]. This identification is supported by two other sets of experiments[17,20] on the same 2H-$NbSe_2$ system. In one[17], the onset of the peak effect was seen to be accompanied by plastic flow (spatially inhomogeneous motion) of a highly defective lattice. In the other case, noise measurements[20] showed that the aforementioned defective moving state is closely related to a defective pinned state, implying that the plastic flow occurs through channels defined by dislocations that persist in the pinned state.

Most of the history dependence in $\chi^{/}$ disappears (near) above the jump seen at $T_p$, i.e., the flux lines system loses order in equilibrium above $T_p$ ( thus it could be well described by a pinned liquid[5,21] or an amorphous lattice, implying the occurrence of FLL melting ). The correlation length $R_c$ of flux lattice presumably becomes[4] of the order of $a_o$ and remains nearly unchanged at higher temperatures. The collapse of the residual pinning at even higher temperature ( $T > T_p$ ) is then caused by the rapid decrease of the pinning interaction $f_p$. The inset of Fig.2(c), comprising plots of $T_{pl}$ and $T_p$ vs H in sample C, further illustrates the role of increased effective pinning at higher fields through the broadening of the peak regime between $T_{pl}$ and $T_p$ and the concomitant shrinkage of the *ordered* state below $T_{pl}$. $T_{pl}$(H) and $T_p$(H) lines may indeed meet at a multicritical point at a low field value.

Above $T_p$, the vortex lattice remains pinned upto at a higher temperature $T_{irr}$, identified in the inset of Fig.1 for sample A. ( Similar data in samples B and C, not highlighted here for brevity. ) Above $T_{irr}$, $\chi^{/}$ becomes positive, due to the differential paramagnetic effect[22] (DPE) : dM/dH > 0, for the reversible part of the M-H curve as diamagnetism decreases with increase in H (dM/dH ~ $\kappa^{-2}$ near $T_c$(H); for $NbSe_2$ [9], $\kappa \sim 10$). $T_{irr}$ occurs near the upper edge of the peak effect regime and is frequency dependent, implying that depending on residual pinning, there is a further (dynamic) crossover from a pinned liquid state to an unpinned vortex liquid state as the viscoelastic relaxation time ( plastic time) decreases with increasing T[5,21].

II . *Effects of disorder at low fields :*

Now we concentrate on the low field regime. We recall that in this regime a reentrant behavior in $T_p$ vs H was observed in the 2H-$NbSe_2$ system[12], consistent with theory[2] and simulations[23]. This regime has been sparingly studied experimentally and little is known about its characteristics and the effect of pinning on it. Fig.3 shows the variation of $\chi^{/}$ with T at low fields for the three 2H-$NbSe_2$ samples under study. For all the samples, the zero-field $\chi^{/}$ curves have also been plotted for easy comparison. In contrast with the high field data - where *increasing* field leads to an effectively larger pinning - in the low field regime the apparent dominance of pinning with *decreasing* field is caused by the weakening of the inter-vortex interaction as the lattice constant $a_0$ approaches the range of interaction, which is comparable to the penetration depth $\lambda$ . At low fields ( for H < 1 kOe ; $a_0 > 1.5 \times 10^3$ A$^o$, indeed comparable to $\lambda$ ( // c ) of $2.3 \times 10^3$ in 2H-$NbSe_2$[9]), one does not see the features observed above, e.g., neither a double peak structure nor jumps spanning the peak regime, etc., are present. Instead, one finds the general trends described below and summarized in the inset of Fig.4 :

(1) The peak regime broadens (i) with increasing pinning ( cf. Figs. 3(a) to 3(c) for the three samples ) and (ii) with decreasing field ( for a given sample ), as indicated by the error bars in the inset of Fig.4. This can be understood to result from an enhanced inhomogeneity in the sizes of Larkin[4] domains within a given sample at a given field and, thence, to a larger spread in the melting temperature of the entire flux line system. This would imply that for a highly disordered solid, the melting phenomenon is a diffuse transition.

(2) For the upper branch of the $T_p$ curve, i.e., FLL melting curve, the $T_p(H)$ boundary shifts down to lower T for more strongly pinned samples ( see, inset of Fig.4 ), showing that a disordered solid ( which is correlated over a smaller volume ) requires less thermal fluctuations to melt it[13].

(3) The reentrant lower branch of the peak effect (melting) line is seen clearly only for sample B and not so for the other two samples (see inset of Fig. 4). The $T_p(H)$ data at low fields in the cleanest sample A shows a pronounced departure away from the upper melting curve well-described by the standard Lindemann melting relation[5] : $B_m = \beta_m (c_l^4/G_i)H_{c2}(0) (1-T/T_c)^2$ with $\beta_m = 5.6$, $G_i = 3 \times 10^{-4}$, $H_{c2}(0) = 4.6$ T [9,12] and a Lindemann number $c_l \sim 0.15$. The observed behavior is in accord with a recent[3] theoretical scenario which shows a similar rapid drop of the melting curve, away from the $H_{c2}(T)$ phase boundary, at low fields (i.e., above the nose or the turnaround of the melting boundary). The reentrant phase boundary may thus be at much lower fields in the very clean crystal A, as proposed theoretically[3] and outside the precision of the present experiment *This implies that, with increasing quenched disorder, the ordered phase is destabilized from above, on the upper curve, by thermal fluctuations and from below on the lower curve, by weaker interaction, thereby shrinking the ordered vortex solid phase region on both branches of the melting curve in the (H,T) phase diagram.*

(4) Finally, the disappearance of the peak effect altogether at lower fields is consistent with the scenario that a transition occurs between an entangled liquid phase of vortices to a pinning dominated disentangled liquid[5] or a glassy phase at the Nelson - Le Doussal[24] line. This line marks the locus ( in H ) where the entanglement length equals the pinning length[5] and scales with $(j_c)^{1/2}$; this accounts well for the increased field values for the disappearance of PE peak in the more impure crystal.

Taking all these results together we construct , in the main panel of Fig.4, a generic phase diagram of a type II superconductor in the presence of both quenched disorder and thermal fluctuations. The nearly defect-free ( elastically deformed ) vortex solid undergoes loss of order in steps. At the $H_{pl}$ line, an elastic-to-plastic transition leads to a plastically deformed solid which melts into a highly viscous pinned liquid at $H_p$. Depending on pinning, both the transitions ( at $T_{pl}$ and at $T_p$ ) can be discontinuous or continuous (within experimental error). At a higher field, $H_{irr}$, a (presumably dynamic) crossover occurs into an unpinned liquid state. $H_{c2}$ marks the final crossover into the normal state. At low fields, $H_{pl}$ and $H_p$ approach each other ( see Fig. 4 ) and a single transition from the vortex solid to a liquid below a multicritical point cannot be ruled out. At still lower H, and particularly near the nose of the peak effect phase boundary, the melting transition broadens significantly, eventually becoming undetectable as a peak effect. This experimentally ascertained phase diagram can be compared with theoretical expectations ( e.g., Fig. 25 of Ref. 5 ) for the mixed state.

It is important to note that, as shown above, the location and nature of these phase transitions and crossovers are explicitly dependent on effective pinning which produces a richness of behavior such as metastability, history dependence and disorder induced transitions not present in the cleanest system. Our results demonstrate that these effects are experimentally accessible and call for more theoretical as well as experimental work.

We acknowledge discussions with G. Blatter, C. Dasgupta, D. Feinberg, V. Geshkenbein, T. Giamarchi, D. Huse, P. Le Doussal and T. V. Ramakrishnan and thank S. Majumdar and N. Trivedi for a critical reading of the manuscript.

@ Corresponding authors: *grover or shobo@tifrvax.tifr.res.in*

1. A. A. Abrikosov, Sov. Phys. JETP **5**, 1174 (1957)
2. D. R. Nelson Phys. Rev. Lett. **60**, 1973 (1988)
3. G. Blatter and D. Geshkenbein, Phys. Rev. Lett. **77**, 4958(1996)
4. A. I. Larkin, Sov. Phys. JETP **31**, 784 (1974) ; A. I. Larkin and Yu. N. Ovchinnikov, J. Low Temp. Phys. **34**, 409 (1979) ; D.S. Fisher, M. P. A. Fisher and D. A. Huse, Phys. Rev. **B 43**, 130 (1991); D. R. Nelson and V. M. Vinokur, Phys. Rev. Lett. **68**, 2398 (1992)
5. G. Blatter et al, Rev. Mod. Phys. **66**, 1125, 1994 and references therein
6. K. H. Fischer and J. A. Hertz , Spin Glasses, Cambridge University Press, Cambridge, U. K., 1991
7. C. P. Bean, Rev. Mod. Phys., **36,** 31 (1964)
8. P. W. Anderson, in Basic Notions in Condensed Matter Physics, Addison - Wesley, New York, U. S. A., 1983 , pp. 162 - 163
9. S. Bhattacharya and M.J. Higgins, Phys. Rev. Lett. **73**, 2617 (1993); M. Higgins and S. Bhattacharya, Physica C **257**, 233 (1996) and references therein
10. A. I. Larkin, M.C. Marchetti and V.M. Vinokur, Phys. Rev. Lett.**75**, 2992 (1995)
11. C. Tang et al, Europhys. Lett. **35**, 597 (1996 ) and references therein
12. K. Ghosh, *et al* Phys. Rev. Lett. **76**, 4600 (1996); S. Ramakrisnan et al, Physica C **256**, 119 (1996)
13. G. I. Menon and C. Dasgupta, Phys. Rev. Lett. **73**, 1023 (1994)
14. W. Henderson *et al*, Phys. Rev. Lett. **77**, 2077 (1996)
15. R. Wördenweber, P.H. Kes and C.C. Tsuei, Phys Rev. **B33**, 3172 (1986) and references therein
16. T. Giamarchi and P. Le Doussal, Phys. Rev. Lett. **72,** 1530 (1994) ; Phys. Rev. **B52**, 1242 (1995).
17. S. Bhattacharya and M.J. Higgins, Phys. Rev. **B43**, 10005 (1994); *ibid.* **B52**, 64 (1995)
18. M. Hellerquist *et al*, Phys. Rev. Lett. **74**, 5114 (1996); M.C. Faleski *et al*, Phys. Rev. B **52**, 12427 (1996) and references therein.
19. M. Gingras and D.A. Huse , Phys. Rev. **B53**, 15193 (1996)
20. R. Merithew *et al*, Phys. Rev. Lett. **77**, 3197 (1996); A.C. Marley *et al*, *ibid.* **74**, 3029 (1995)
21. V. M. Vinokur *et al*., Phys. Rev. Lett. 65, 259 (1990)
22. R.A. Hein and R.A. Falge Jr., Phys. Rev. **123**, 407 (1961)
23. M. I. J. Probert and A. I. M. Rae, Phys. Rev. Lett **75**, 1835 (1995)
24. D.R. Nelson and P. Le Doussal, Phys. Rev. B42, 10113 (1990) ; D.R. Nelson, in The Vortex State, pp 41-61, N. Bontemps et al (eds.), Kluwer Academic Publishers, The Netherlands, 1994.

**Figure Captions :**

**Figure 1 :**Real part ($\chi^{/}$**)** of the ac susceptibility response [$h_{ac}$ = 1.0 Oe (r.m.s) and f = 211 Hz] in our purest (sample A) crystal of 2H-$NbSe_2$ in ( nominally ) zero bias field and in $H_{dc}$ of 4 kOe, respectively. The $\chi^{/}$ curve in $H_{dc}$ = 4kOe shows a negative *peak* at $T_p$ (i.e., peak effect (PE)), prior to $T_c$(H), whose sharpness exceeds that of the zero field superconducting transition. The inset reveals a small *paramagnetic peak* due to differential paramagnetic effect (DPE) in between the irreversibility temperature $T_{irr}$(H) and $T_c$(H).

**Figure 2 :** $\chi^{/}$(T) curves for field cooled (FC) and zero field cooled (ZFC) states at selected fields in three crystals of 2H-$NbSe_2$. ZFC data points are marked in red whereas continuous lines (in blue) pass through closely spaced FC data points (not marked). In cleanest sample A (panel (a)), FC and ZFC data points overlap in all fields, whereas in samples B and C (panels (b) and (c)), the differences between FC and ZFC data can be seen in H = 15 kOe and 5 kOe, respectively. $T_{pl}$ and $T_p$, respectively, mark the temperatures of the onset of anomalous change and the maximum of diamagnetic screening response in $\chi^{/}$, in each of the figures. The insets in Fig. 2(a) and Fig. 2(b) show FC and ZFC data recorded at H = 2 kOe in sample A and at H = 5 kOe and 2 kOe in sample B, respectively ( see text for description). The inset in Fig.2 (c) shows variation of $T_{pl}$ and $T_p$ with H in sample C.

**Figure 3 :** $\chi^{/}$(T) curves at different H ( $\leq$ 1 kOe ) in crystals A, B and C of 2H-$NbSe_2$ . The zero field $\chi^{/}$ curve is sharpest in A and broadest in C, thereby showing the progressive decline in the purity. Also note, the peak regime broadens and $T_p$(H) (for a fixed H of 300 Oe ) shifts to progressively lower reduced temperature value on going from A to C. In sample A, the PE peak remains very sharp even at H = 30 Oe, whereas in sample B, it broadens rapidly below 300 Oe. In sample C, a distinct PE peak is not identifiable below 300 Oe.

**Figure 4 :** A schematic plot of magnetic phase diagram in a weakly pinned type - II superconductor ( here, for example, crystal B ). Vortex ( mixed ) state is sandwiched between the Meissner state ( below lower critical field $H_{c1}$ line) and the normal metallic state ( above upper critical field $H_{c2}$ ). Different phases of vortex lines have been identified on the basis of experiments in 2H-$NbSe_2$ system. $H_{pl}$, $H_{melt}$, $H_{irr}$ and $H_{c2}$ lines correspond to $T_{pl}$(H), $T_p$(H), $T_{irr}$(H) and $T_c$(H), respectively, in Figs. 1 to 3. Nelson - Le Doussal[24] line has been drawn following the schematic plot in Fig. 1 of Ref. 24 and made to coincide with the disappearance of PE. The inset shows the loci of peak effect temperatures $T_p$(H) at low fields (H $\leq$ 1kOe; // c) in crystals A, B and C of 2H-$NbSe_2$. Experimentally determined[12] $H_{c1}$(T) curve for H//c has also been drawn for the sake of completeness.